\documentclass[cameraready]{Interspeech}   %
\makeatletter
\newcommand{\email}[1]{\gdef\@email{#1}}
\makeatother

\title{End-to-End Direction-Aware Keyword Spotting with Spatial Priors in Noisy Environments}

\author{Rui}{Wang}
\author{Zhifei}{Zhang}
\author{Yu}{Gao$^{*}$}
\author{Xiaofeng}{Mou}
\author{Yi}{Xu}
\address{AI Research Center, Midea Group (Shanghai) Co., Ltd., China}
\email{\{wangrui302, zhangzf148, gaoyu11, mouxf, xuyi42\}@midea.com}


\usepackage{comment} \usepackage{amsmath,graphicx,hyperref} \usepackage{amsfonts} \usepackage{color} \usepackage{amsmath} \usepackage{bm} \usepackage{bbding} \usepackage{graphicx} \usepackage{slashbox} \usepackage{makecell} \usepackage{latexsym} \usepackage{hyperref} \usepackage{cite} \usepackage{numprint}

\begin{document}
\maketitle

\begingroup
\renewcommand{\thefootnote}{\fnsymbol{footnote}}
\footnotetext[1]{Corresponding author.}
\endgroup
\begin{abstract}
Keyword spotting (KWS) is crucial for many speech-driven applications, but robust KWS in noisy environments remains challenging. Conventional systems often rely on single-channel inputs and a cascaded pipeline separating front-end enhancement from KWS. This precludes joint optimization, inherently limiting performance. We present an end-to-end multi-channel KWS framework that exploits spatial cues to improve noise robustness. A spatial encoder learns inter-channel features, while a spatial embedding injects directional priors; the fused representation is processed by a streaming backbone. Experiments in simulated noisy conditions across multiple signal-to-noise ratios (SNRs) show that spatial modeling and directional priors each yield clear gains over baselines, with their combination achieving the best results. These findings validate end-to-end multi-channel spatial modeling, indicating strong potential for the target-speaker-aware detection in complex acoustic scenarios.
\end{abstract}

\section{Introduction}

\label{sec:intro}

Keyword spotting (KWS), also known as wake word detection (WWD), aims to detect a predetermined wake word from continuous speech audio streams. KWS is widely adopted as the entry point to voice interfaces and is widely deployed in many applications, particularly commercial assistants such as Amazon Alexa, Apple Siri, Google Assistant, and Microsoft Cortana \cite{app2018}\cite{google}. 

Real-world applications of voice interfaces including KWS and ASR are often required to respond to specific target speaker's command. Meanwhile, the acoustic environment in real scenes is also more complex, where background noise, reverberation, and overlapping speech degrade performance. A common remedy is a cascaded pipeline that inserts a front-end enhancement/beamforming stage before the acoustic model~\cite{cascaded1,cascaded2,cascaded3,cascaded3.1,cascaded4, cascaded4.2, cascaded5}. With the wider deployment of microphone arrays, multi-channel processing has become increasingly important; studies show that leveraging spatial information about the target source can substantially improve separation/extraction when multi-channel signals are available~\cite{doa1,doa2,doa3}. Yet in traditional cascades, decoupling the front-end from the detector hinders joint optimization and risks objective mismatch~\cite{front_end_dis0,front_end_dis1,front_end_dis2}. Even with extensive data augmentation (e.g., additive noise, RIR convolution), robustness under noise and interference remains challenging~\cite{augmentation}.

Recent advances on deep learning have shifted KWS toward end-to-end (E2E) modeling, which jointly learns feature extraction and detection~\cite{e2e1, overview}. Toolkits such as \emph{WeKws} provide standardized pipelines and practical streaming recipes ~\cite{wekws}. Nevertheless, effective KWS toward target speaker in complex acoustics conditions like noisy environment remains challenging. Two main gaps remain in these E2E methods. First, many E2E systems assume single-channel inputs or treat multi-channel signals by simple channel stacking, leaving spatial cues underutilized and lacking a learnable spatial module~\cite{front_end_dis1}. Second, directional knowledge is rarely modeled explicitly, so target-speaker awareness in multi-speaker conditions is limited. Consequently, noise robustness and interference suppression are still constrained in realistic deployments. These gaps call for an end-to-end framework that incorporates spatial modeling within the recognizer and allows the injection of direction-aware priors. 

In this paper, we present an end-to-end direction-aware framework for multi-channel KWS in noisy environments. The core is an end-to-end integration of (i) a spatial encoder for feature extraction, (ii) a spatial embedding that injects direction-aware priors derived from spatial cues, and (iii) a streaming KWS model. The spatial encoder learns inter-channel features from multi-microphone signals, while the spatial prior is associated with the multi-channel features, which enhance the KWS for target speaker in noisy environments. Compared with conventional single-channel methods and cascaded systems with an independent front-end enhancement module, the experimental results demonstrate clear advantages of the proposed unified architecture. The remainder of this paper is organized as follows. Section 2 details the proposed framework, including architecture and pipeline; Section 3 describes the experimental evaluations; Section 4 concludes the work.

\begin{figure*}[th]
\begin{center}
    \includegraphics[width=0.86\textwidth]{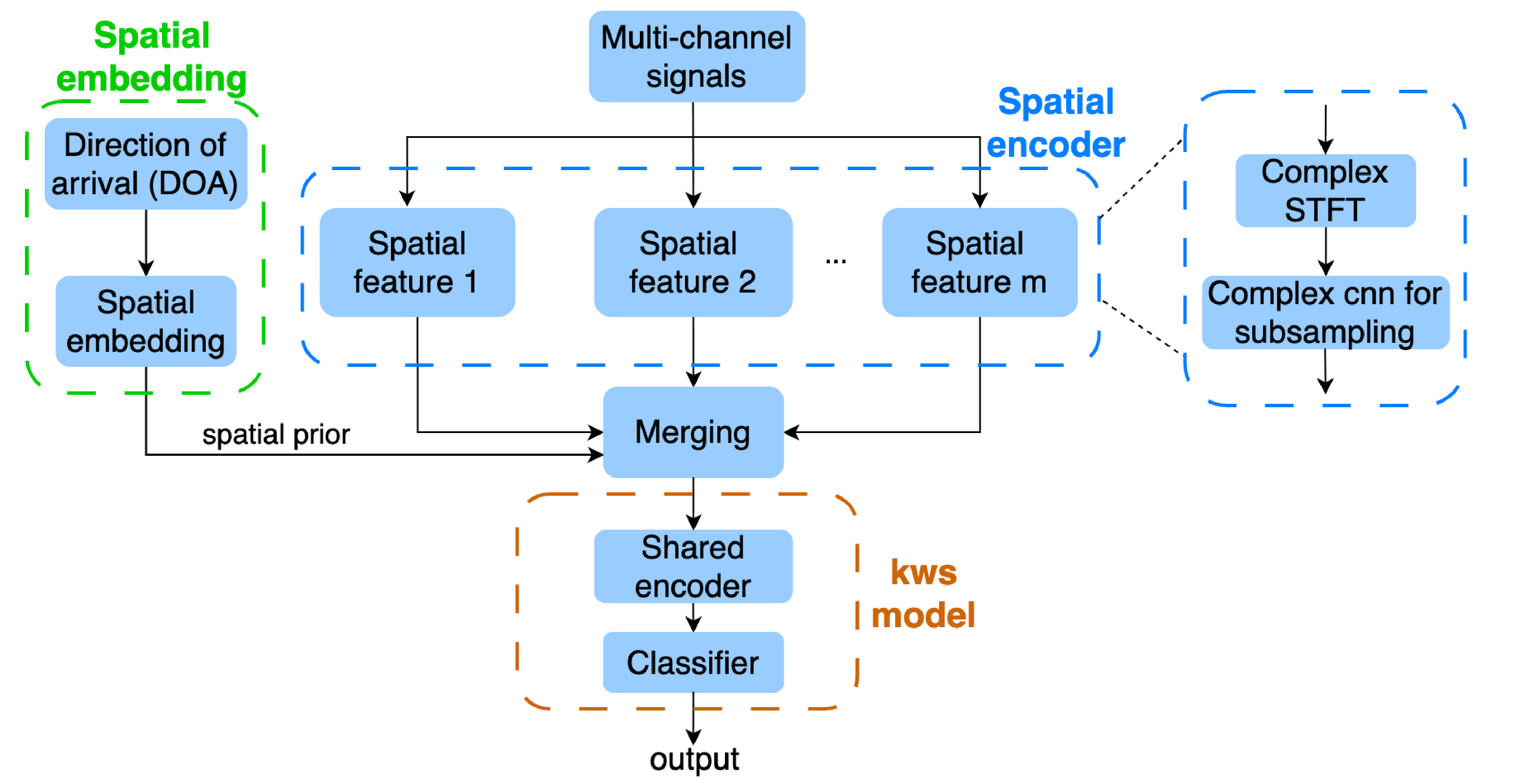}
\end{center}
\caption{\rm Structure of the proposed KWS framework.}
\label{fig:frame}
\end{figure*}

\section{End-to-end direction-aware KWS}
\label{sec:format}
\subsection{Overview}
An end-to-end KWS system typically contains three components: (i) a feature extraction module as the preprocessing (e.g., waveform or spectral features), (ii) an acoustic encoder (backbone) that maps frames/chunks to high-level representations, and (iii) a classifier that converts posterior trajectories into keyword triggers. Most existing end-to-end pipelines are single-channel and learn to detect predefined commands from continuous speech using alignment-free pooling objectives or sequence criteria (e.g., CTC)\cite{ctc1}\cite{ctc2}, but generally do not model spatial cues explicitly. 

\subsection{Proposed E2E spatial-aware framework}
\label{proposed}
To address the above limitations, we propose a KWS framework as an end-to-end, direction-aware integration that jointly learns (i) a spatial encoder from multi-channel signals, (ii) a spatial embedding that injects spatial priors, and (iii) a streaming KWS model. The first two provide complementary conditioning—audio-based spatial acoustics features and label-based directional cues—while the entire network is optimized toward the final KWS objective under a standard loss. The overall framework with the data flowing is shown as Fig. \ref{fig:frame}. The following part of this section will describe the details of each module of the proposed framework and the pipeline in training and inference. 

\subsubsection{Data preparation and input presentation}

We use a multi-channel system in our implementation. Multi-channel waveforms are converted to complex spectral features in the time-frequency (TF) domain, denoted as $X_m(f,t)$, where $m=$ represents the number of channels. These complex spectral features maintain inter-channel phase and magnitude relations. As a result, they carry spatial cues that are useful for downstream modules. Later sections will provide details on data generation and feature configuration. 

\subsubsection{Spatial encoder} 
We operate directly on the multi-channel complex spectral features $X_m(f,t)$ with a two-stage Conv2D subsampler: a complex 2-D convolution (stride along time–frequency) followed by ReLU, then a lightweight real Conv2D with stride. This reduces the temporal resolution by $ (e.g., T \rightarrow T^\prime)$. The spatial encoder $\mathrm{Encoder}_{\mathrm{sp}}$ outputs a time-aligned feature sequence. 

\begin{equation}
\qquad H \in \mathbb{R}^{B\times T'\times d} = \mathrm{Encoder}_{\mathrm{sp}} (X_m(f,t))
\label{eq:emb_min1}
\end{equation}

 \noindent where $B$ represents the batch size, $T'$ is the number of time steps after subsampling, and $d$ is the feature dimension of the encoder output. The output of the spatial encoder preserves inter-channel cues—IPD/ILD-like (inter-channel phase and level differences) without explicit beam synthesis, and is trained end-to-end with the KWS objective.

 \subsubsection{Spatial embedding}
We assume that the target’s direction-of-arrival (DOA) is known during training and evaluation; in deployment, the same interface can be fed by any DOA estimation method. We map a discrete directional label using a lightweight embedding network $\mathrm{Emb}$, which converts the DOA-derived label $\theta$ into a compact prior 

\begin{equation}
e_{\theta} = \operatorname{Emb}(\theta), \quad e_{\theta}\in\mathbb{R}^{d}
\label{eq:emb_min2}
\end{equation}

\noindent where $\theta \in \{1,\dots,K\}$ and $K$ represents the zone index in the acoustics space. The DOA label $\theta$ is discretized into $K$ angular zones covering the array’s field of view (FOV). Here, $K$ controls the spatial resolution of the prior, with an approximate bin width of $FOV/K$. In the spatial embedding network, a two-layer MLP with ReLU nonlinearity refines the representation, followed by dropout and layer normalization to stabilize conditioning. We adopt linear feature merging of the output multi-channel features produced by the spatial encoder with the generated spatial prior as 

\begin{equation}
\tilde H \;=\; H \;+\; e_\theta
\label{eq:emb_min3}
\end{equation}

This fusion biases the model toward the target direction while preserving acoustic evidence. The fused sequence $\tilde H$ is then processed by the streaming backbone described next.

\subsubsection{KWS module}
We adopt the multi-scale depthwise temporal convolution (MDTC) backbone \cite{mdtc} as the shared encoder. MDTC stacks causal depthwise temporal-convolution (DTC) blocks with varying dilations to capture multi-scale temporal context under streaming constraints, effectively enlarging the receptive field without increasing parameter overhead. The encoder takes the fused sequence $\tilde H$ as input and produces a unified representation for the classifier heads. Inference runs in a streaming manner with causal convolutions and cached left context; no future frames are used.

To support multiple wake words, we attach independent binary classifiers after the shared encoder—one single-node sigmoid head per keyword—while the encoder parameters are shared across all keywords. This setup follows standard practice in prior KWS systems \cite{wekws}, keeping the backbone compact and allowing each keyword to be scored and triggered independently. In practice, this design also makes it easy to add or retire keywords by training only the corresponding small head while reusing the same shared encoder. At run time, the heads produce posteriors for their respective keywords; simple posterior smoothing and per-keyword thresholds are used to emit stable triggers. 

\section{EXPERIMENT EVALUATION}
\label{sec:pagestyle}

We carried out experimental evaluations on our proposed method and some baselines. In this section, we will introduce our setting and the evaluation results. 

\subsection{Training and testing setup}

\subsubsection{Datasets}

We used the Google Speech Commands v1 (GSC v1) \cite{gsc} corpus as the source of clean speech and text labels for training. The GSC v1 corpus consists of \numprint{64721} one-second-long recordings of 30 words by \numprint{1881} different speakers. We followed the setup of train/validation/test split in \cite{wekws}, in which 10 of 30 words are set as the keywords.

\subsubsection{Spatial signal simulation and DOA labels}
To obtain spatialized signals with DOA labels, we synthesized multi-channel recordings using gpuRIR \cite{gpuRIR}.\footnote{\url{https://github.com/DavidDiazGuerra/gpuRIR}} Because GSC v1 is single-channel, each utterance was treated as a point source and spatialized by convolving it with room impulse responses (RIRs) generated for specific array geometries. The acoustic environments were consistent across all configurations: rooms were randomly sampled with lengths of 3–8 m, widths of 3–5 m, heights of 2.5–4 m, and $RT_{60}$ of 0.05–0.8 s, with the source–array distance drawn from 0.5–5 m. 

We simulated two microphone configurations. For the dual-channel setup, a linear array with 3 cm spacing was used. The target DOA $\theta$ was constrained to the front hemisphere ($0^{\circ}$--$180^{\circ}$) and discretized into six $30^{\circ}$ space zones (yielding labels 1--6). For the extended three-channel setup, the spatial coverage was expanded to a full $360^{\circ}$ azimuth, and the target DOA was correspondingly discretized into twelve uniform $30^{\circ}$ zones (labels 1--12).  In both configurations, an additional label 0 is reserved to denote the ``no spatial prior'' condition for baseline training and testing. Thus, each rendered utterance comes with a multi-channel waveform and a discrete zone label $\theta$.

\subsubsection{Noise rendering}
Background noise was taken from the DEMAND dataset \cite{demand}, which contains 17 categories of noise. We used 12 categories for training and another 5 categories for testing. Noise was spatialized by convolving with independently sampled multi-channel RIRs (from the same simulator/room pool) and then mixed with the target at a signal-to-noise ratio (SNR) uniformly sampled in $[0, 10]$ dB, during training. 

\subsubsection{Objective}
The loss function for training was cross-entropy over $C$ classes (keywords plus a non-keyword class). We averaged the loss over valid frames after flattening the time axis; per-frame accuracy was also reported as a training diagnostic.

\subsection{Evaluation of KWS performance under noisy conditions}
\subsubsection{Test rendering}
Testing multi-channel signals were generated with the same simulator, array, and DOA discretization as training, but with fixed SNR conditions of 0, 5, and 10 dB. For each SNR, we rendered an independent test set of \numprint{6835} test samples. Unless otherwise noted, spatial information used for priors in our proposed system comes from the rendering metadata of generated RIRs. 

\subsection{Systems under comparison}

\subsubsection{Single-channel baseline}
\label{baseline1}
We used the original WeKws implementation as a single-channel KWS baseline with the same Mel-filter bank (Fbank) feature setting as in \cite{wekws}. For the spatialized two-channel signals, we fed channel 1 only to the backend, so no spatial information is used.

\subsubsection{Enhanced cascaded baseline}
We inserted a generalized sidelobe canceller (GSC) \cite{gsc_b} beamformer in front of the WeKws backend to form a cascaded system. The multi-channel input was first enhanced using an independent GSC method; then one beam output was passed to the same WeKws pipeline as in Sec. \ref{baseline1}. This baseline keeps access to spatial information via the beamformer but decouples it from the recognizer. We used it to assess how a cascaded design with spatial enhancement compares with our end-to-end approach under the same spatial conditions.

\subsubsection{Dual-channel E2E system without prior}
Our end-to-end dual-channel system without spatial prior (\textit{2ch E2E without prior}): during both training and test, the DOA label was fixed to $\theta=0$ (no-prior token). This baseline relies entirely on the internal convolutional layers of the spatial encoder to implicitly extract inter-channel phase differences across the two sensors. Other settings were unchanged. 

\subsubsection{Proposed dual-channel spatial E2E system}
The same dual-channel end-to-end system with spatial priors (\textit{2ch Proposed spatial E2E}) as we presented in Sec. \ref{proposed}. DOA labels (1–6) are embedded and fused with the encoder features as described in Sec. \ref{proposed}.


\subsubsection{Extension to three-channel system}
To rigorously evaluate the scalability of the spatial encoder and the impact of higher angular resolution, we extend the end-to-end systems from a dual-channel to a three-channel system. This extension introduces modifications to the input dimensions and yields two additional configurations.

First, for the \textit{\textbf{3ch E2E without prior}}, the acoustic front-end is adapted to ingest three-channel stacked audio inputs. The external DOA label is strictly fixed to $\theta=0$ (the no-prior token) during both the training and testing phases. 

Second, for the \textit{\textbf{3ch Proposed spatial E2E}}, the system leverages the denser spatial sampling provided by the extra microphone. The target DOA space is mapped to the finer 12-zone discrete labels (1--12), representing a full $360^{\circ}$ azimuth. The spatial encoder is correspondingly expanded to accommodate this larger vocabulary of directional indices. Except for the input channel dimensions and the spatial embedding vocabulary size, all other network hyper-parameters and training criteria remain identical to the dual-channel setups.

\begin{table}[t]
\begin{center}

\caption{Average accuracy of KWS under different SNRs.}
\label{tab:evaluation}

\begin{tabular}{c|cc}
\hline 
Method &  \multicolumn{2}{c}{SNR = 0 dB} \\ 
       &  parameters & accuracy (\%) \\ \hline
Single-channel baseline \cite{wekws} &  164k& 69.86  \\ \hline
Enhanced cascaded baseline & 164k &72.19   \\ \hline
2ch E2E without prior &  279k & 76.92 \\ \hline
3ch E2E without prior &  279k & 80.23 \\ \hline
\textbf{2ch Proposed spatial E2E} &  \textbf{279k}&  \textbf{77.67}  \\ \hline 
\textbf{3ch Proposed spatial E2E} &  \textbf{279k}&  \textbf{79.39}  \\ \hline 
\end{tabular}  \\

\begin{tabular}{c|cc}
\hline 
Method &  \multicolumn{2}{c}{SNR = 5 dB}  \\ 
       &  parameters & accuracy(\%)   \\ \hline
Single-channel baseline \cite{wekws} & 
164k& 75.89  \\ \hline
Enhanced cascaded baseline & 164k & 78.19   \\ \hline
2ch E2E without prior &  279k & 84.26  \\ \hline
3ch E2E without prior &  279k & 85.85  \\ \hline
\textbf{2ch Proposed spatial E2E} & \textbf{279k}& \textbf{85.25}  \\ \hline 
\textbf{3ch Proposed spatial E2E} & \textbf{279k}&  \textbf{85.22}  \\ \hline
\end{tabular}  \\

\begin{tabular}{c|cc}
\hline 
Method &  \multicolumn{2}{c}{SNR = 10 dB}  \\ 
      &  parameters & accuracy (\%)  \\ \hline
Single-channel baseline \cite{wekws}& 164k & 81.68   \\ \hline
Enhanced cascaded baseline & 164k & 82.87   \\ \hline
2ch E2E without prior & 279k& 88.57\\ \hline
3ch E2E without prior & 279k& 89.40\\ \hline
\textbf{2ch Proposed spatial E2E} & \textbf{279k}& \textbf{89.42} \\ \hline
\textbf{3ch Proposed spatial E2E} & \textbf{279k}& \textbf{89.61} \\ \hline
\end{tabular}

\end{center}
\end{table}

\section{Results and Discussion}

\subsection{Overall Performance and Baseline Comparison}
Table 1 summarizes the results and model parameters. Our dual-channel end-to-end KWS with a directional prior ranks first among 2-channel setups across all SNRs. At 0 dB, it achieves 77.67\% accuracy, outperforming the single-channel WeKws baseline by 11.18\% relatively and the beamformer cascade by 5.48\% absolutely. 

With a comparable parameter budget, end-to-end systems consistently outperform non-end-to-end baselines. Although the cascaded system improves upon the single-channel baseline, it remains limited by the separation of front-end and back-end, where processing artifacts introduced during enhancement cannot be jointly optimized with the downstream recognizer. This confirms that joint spatial and acoustic modeling in a unified end-to-end framework is more effective than separated pipelines.

\subsection{Impact of Spatial Priors}
Across both dual-channel and three-channel configurations, the performance margin between models with spatial priors and their no-prior baselines is relatively small. This is primarily because the current evaluation lacks distinct interfering speakers. Without strong directional interference, the multi-channel acoustic features alone are highly informative for robust keyword detection under background noise. Consequently, the no-prior models remain strong, relying on learned spatial representations, dynamically adapting to spatial cues and achieving strong performance independently.

Nevertheless, the directional prior affects the models differently depending on the spatial task complexity. In the dual-channel setup, discretizing the $180^{\circ}$ space into 6 zones yields a coarse prior that is generally less sensitive to DOA mismatch. This broad cue provides a small but consistent gain over the no-prior baseline across SNR conditions, even when noise degrades the acoustic cues.

Conversely, the three-channel system expands the prior to a 12-zone, $360^{\circ}$ full-azimuth layout, doubling the spatial vocabulary. At lower SNRs (0 and 5 dB), conditioning this strict, high-resolution prior on noise-smeared features increases the risk of feature-prior mismatch. Thus, the no-prior baseline, retaining its data-driven flexibility, slightly surpasses the proposed model. However, at 10 dB SNR, the clean acoustic features reliably align with the complex 12-zone prior. Here, the spatial prior acts as a stronger directional bias, allowing the proposed 3-channel model to achieve the overall highest accuracy (89.61\%). This suggests that prior granularity should be matched to the reliability of spatial cues under varying noise intensities, which we will further analyze with DOA-error perturbations in more complex acoustic scenes.

\section{Conclusion}
\label{sec:prior}
This paper presented an end-to-end multi-channel KWS framework that unifies a learnable spatial encoder, a direction-aware embedding, and a streaming shared backbone. By explicitly leveraging inter-channel cues and injecting a discrete spatial prior, the system achieves stronger noise robustness than a single-channel WeKws baseline and a beamformer-enhanced cascaded pipeline at a similar parameter budget. Evaluations across dual-channel and three-channel systems demonstrate that the model remains robust without priors, yet benefits when they are present. Our analysis further highlights a practical trade-off: while high-resolution priors act as precise spatial filters in clear conditions, lower-complexity priors offer better fault tolerance under heavy noise. These findings provide valuable architectural guidelines for deploying reliable voice-controlled interfaces in diverse, real-world acoustic environments.

Our method is modular and extensible. In future work, a trainable DOA estimator can be integrated into a multi-task framework to dynamically provide zone-wise or continuous directions to condition the spatial embedding. We also plan to incorporate an enhancement front-end alongside the spatial encoder, yielding a joint localize--enhance--wake pipeline. This joint optimization is expected to further mitigate feature-prior mismatches in highly reverberant conditions. Richer prior designs, such as probabilistic spatial embeddings, will be explored to improve robustness against DOA mismatch. Finally, we will investigate on-device, weakly supervised fine-tuning to deploy this unified system under strict streaming constraints on latency and memory.

\bibliographystyle{IEEEtran}
\bibliography{mybib}

\end{document}